\documentclass{svproc}
\usepackage{url}

\usepackage[utf8]{inputenc}
\usepackage[square,sort&compress,comma,numbers, sectionbib]{natbib}
\usepackage{ amssymb }
\usepackage{amsmath}
\usepackage{mathrsfs}  

\usepackage{graphicx}
\usepackage{subcaption}

\usepackage[dvipsnames]{xcolor}
\usepackage[pdftex, 
colorlinks,
pdfpagelabels,
pdfstartview = FitH,
bookmarksopen = true,
bookmarksopenlevel = 2,
bookmarksnumbered = true,
linkcolor = blue!80!black,
plainpages = false,
hypertexnames = false,
urlcolor = blue!80!black,
citecolor = blue!80!black] {hyperref}

\usepackage[english]{babel}

\begin{document}
\mainmatter              
\title{Impact of restricted spin-ranges in the \\ Oslo  Method: The example of (d,p)$^{240}\mathrm{Pu}$}
\titlerunning{Restricted spin-ranges in the Oslo Method}  
%
\author{
	F. Zeiser\inst{1}$^{,\dagger}$
	\and
	G. Potel\inst{2}
	\and
	G.M. Tveten\inst{1}
	\and
	A.C. Larsen\inst{1}
\and
M. Guttormsen\inst{1}
\and
T.A. Laplace\inst{3}
\and
S. Siem\inst{1}
\and
D.L. Bleuel\inst{4}
\and
B.L. Goldblum\inst{3}
\and
L.A. Bernstein\inst{3,5}
\and
F.L. Bello Garrote\inst{1}
\and
L. Crespo Campo\inst{1}
\and
T.K. Eriksen\inst{1}
\and
A. Görgen\inst{1}
\and
K. Hadynska-Klek\inst{1}
\and
J.E. Midtbø\inst{1}
\and
T. Renstrøm\inst{1}
\and
E. Sahin\inst{1}
\and
T. Tornyi\inst{1}
\and
A. Voinov\inst{6}
\and
M. Wiedeking\inst{7}
}
\authorrunning{F. Zeiser et al.} 
%
\tocauthor{F. Zeiser, et al.}
\institute{
	Department of Physics, University of Oslo, P.O. Box 1048 Blindern, N-0316 Oslo, Norway
	\and Facility for Rare Isotope Beams, Michigan State University, East Lansing, MI 48824, USA
	\and Department of Nuclear Engineering, University of California, Berkeley, 94720, USA
	\and Lawrence Livermore National Laboratory, Livermore, CA 94551, USA
	\and Lawrence Berkeley National Laboratory, Berkeley, California 94720, USA
	\and Department of Physics and Astronomy, Ohio University, Athens, Ohio 45701, USA
	\and iThemba LABS, P.O. Box 722, Somerset West 7129, South Africa\\
	$^{\dagger}$ Corresponding author:  \email{fabio.zeiser@fys.uio.no}}

\maketitle              

\begin{abstract}
In this paper we present the first systematic analysis of the impact of the populated vs. intrinsic spin distribution on the nuclear level density and $\gamma$-ray strength function retrieved through the Oslo Method. We illustrate the effect of the spin distribution on the recently performed $^{239}\mathrm{Pu}$(d,p$\gamma$)$^{240}\mathrm{Pu}$ experiment using a 12 MeV deuteron beam performed at the Oslo Cyclotron Lab. In the analysis we couple state-of-the-art calculations for the populated spin-distributions with the Monte-Carlo nuclear decay code RAINIER to compare Oslo Method results to the known input. 
We find that good knowledge of the populated spin distribution is crucial and show that the populated distribution has a significant impact on the extracted nuclear level density and $\gamma$-ray strength function for the $^{239}\mathrm{Pu}$(d,p$\gamma$)$^{240}\mathrm{Pu}$ case.
\end{abstract}
\section{Introduction}
Proper knowledge of neutron-induced cross-sections from thermal energies to several MeV is important for many physical applications. However, the lack of a mono-energetic neutron source in the full energy range hampers direct cross-section measurements. The short half-lives of many isotopes of astrophysical interest make it impossible to create targets for direct measurements using neutron beams. In these cases, calculations within the statistical framework can provide an alternative approach to obtain (n,$x$) cross-sections. These rely essentially on precise measurements of nuclear level densities (NLD) and $\gamma$-ray strength functions ($\gamma$SF)~\cite{Ullmann2017}.

The Oslo Method~\cite{Schiller2000, Larsen2011} can be used to analyze particle-$\gamma$ coincidence spectra from transfer reactions to simultaneously extract NLDs and $\gamma$SFs. In a campaign to study actinides the method has been applied to the compound nuclei $^{231-233}\mathrm{Th}$, $^{232,233}\mathrm{Pa}$, $^{237-239}\mathrm{U}$, $^{238}\mathrm{Np}$~\cite{Guttormsen2012, Guttormsen2013,Guttormsen2014, Tornyi2014a} and  $^{243}\mathrm{Pu}$~\cite{Laplace2015} using different light-ion reactions. The extracted $\gamma$SFs show a significant enhancement between about 2 and 4 MeV, which is consistent with the location~\cite{Heyde2010} of a low energy orbital M1 scissors resonance (SR).

Larsen \textit{et al.}~\cite{Larsen2011} have shown that the population of a limited spin range may lead to distortions of the $\gamma$SF. This has been observed in some of the previous studies on actinides~\cite{Guttormsen2012,  Guttormsen2013, Guttormsen2014, Tornyi2014a, Laplace2015} due to the low-spin transfer using the $(d,p)$ reaction mechanism, where an ad hoc procedure for the correction was adopted. In this proceeding, we focus on the first systematic analysis of the impact on the Oslo Method results for a realistic spin-parity population for the (d,p)$^{240}\mathrm{Pu}$ reaction.

\section{Experimental Setup and Data Analysis}
\label{sec:exp}

The $(\mathrm{d},\mathrm{p})^{240}\mathrm{Pu}$ experiment was conducted using a 12 MeV deuteron beam at the Oslo Cyclotron Laboratory (OCL). The $0.4\,\mathrm{mg}/\mathrm{cm}^2$ thick $^{239}\mathrm{Pu}$ target was purified using an anion-exchange resin column procedure~\cite{Henderson2011} prior to electroplating on a $1.9\,\mathrm{mg}/\mathrm{cm}^2$ beryllium backing.

The outgoing charged particles were detected with the SiRi particle telescope~\cite{Guttormsen2011}. SiRi consists of 64 silicon particle telescopes with a thickness of 130 $\mu$m for the front ($\Delta E$) and 1550 $\mu$m for the back ($E$) detectors, and was placed at backwards angles ($126^\circ$ to $140^\circ$). The CACTUS array~\cite{Guttormsen1990} measured coincident $\gamma$ rays and was composed of 26 lead collimated $5^{\prime\prime} \times 5^{\prime\prime}$ NaI(Tl) crystals with a total efficiency of 14.1(2)$\%$ at $E_\gamma= 1.33\,\mathrm{MeV}$. Additionally, four Parallel Plate Avalanche Counters (PPAC)~\cite{Tornyi2014} were used to detect fission events.

The reaction kinematics allowed for selection of (d,p) events and conversion of the detected proton energy to the excitation energy $E_\mathrm{x}$ of the compound nucleus $^{240}\mathrm{Pu}$. Prompt  $\gamma$ rays were selected from a $\pm 14$\,ns wide time-window with background correction applied. The $\gamma$-ray spectra were unfolded following the procedure of~\cite{Guttormsen1996}, using response functions~\cite{Campo2016} that were updated in 2012.

Next, an iterative subtraction technique~\cite{Guttormsen1987} was applied to obtain the primary  $\gamma$ rays $P(E_\mathrm{x},E_\gamma)$ (also called first-generation $\gamma$ rays) for each $E_x$ bin from the initial spectra, which include all $\gamma$ rays of the decay cascades. Here we relied on the assumption that the (d,p)-reaction will populate a similar spin-parity distribution for the levels in an $E_x$ bin $i$ as would be populated from $\gamma$-decay from a higher excitation energy bin $j$. Consequently, by subtracting the $\gamma$-ray spectra from bins with lower excitation energy, only the primary  $\gamma$ rays remain. 


\section{Extraction of NLD and $\gamma$SF}

For $\gamma$ rays  emitted in the statistical regime (i.e., high level density) we can determine the NLD at the excitation energy of the final state, $\rho(E_\mathrm{x,f})$, and the $\gamma$-ray transmission coefficient, $\mathscr{T}(E_\gamma)$~\cite{Schiller2000}:
\begin{equation}
P(E_\mathrm{x,i},E_\gamma) \propto \rho(E_\mathrm{x,f}) \mathscr{T}(E_\gamma),
\label{eq:LevelDensityTransmissionCoefficient}
\end{equation}
up to a transformation with the parameters $A$, $B$ and $\alpha$,
\begin{align}
	\tilde{\rho} (E_i - E_\gamma) & = A \exp[\alpha \,(E_i - E_\gamma) ]\,  \rho(E_i - E_\gamma), \label{eq:rho_trans}\\
	\tilde{\mathscr{T}}(E_\gamma) &= B \exp[ \alpha \, E_\gamma ]\, \mathscr{T}(E_\gamma). \label{eq:t_trans}
\end{align}
To select the $\gamma$ decay channel, only excitation energies $E_\mathrm{x}$ below the neutron separation energy ($S_n = 6.534\,\mathrm{MeV}$~\cite{Singh2008}) must be considered. In this experiment, we applied more stringent constrains due to the onset of sub-barrier fission events at about 4.5 MeV ~\cite{Back1974, Hunyadi2001}. A more detailed analysis of the prompt fission  $\gamma$ rays can be found in~\cite{Rose2017}. The final extraction regions were $E_{\gamma}^\mathrm{min}=1.2\,\mathrm{MeV}$, $E_{\mathrm{x}}^\mathrm{min}=2.4\,\mathrm{MeV}$, $E_{\mathrm{x}}^\mathrm{max}=4.0\,\mathrm{MeV}$. It remained then to find the transformation parameters corresponding to the correct physical solution.

\begin{figure}[t]
\centering
	\begin{subfigure}[b]{0.49\textwidth}
	\includegraphics[width=\linewidth]{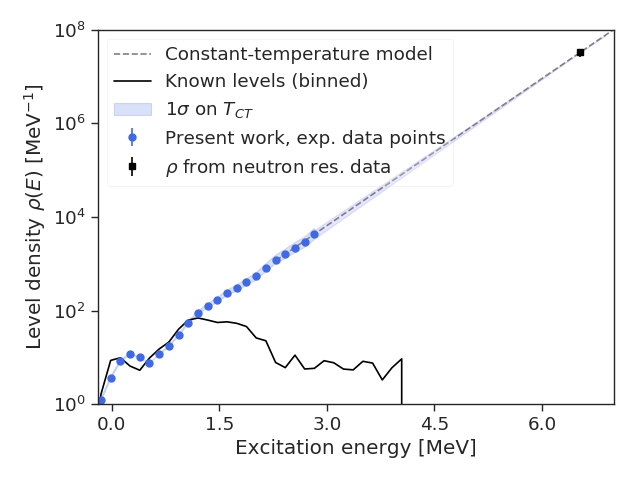}
	\caption{}
	\end{subfigure}
	\begin{subfigure}[b]{0.49\textwidth}
		\includegraphics[width=\linewidth]{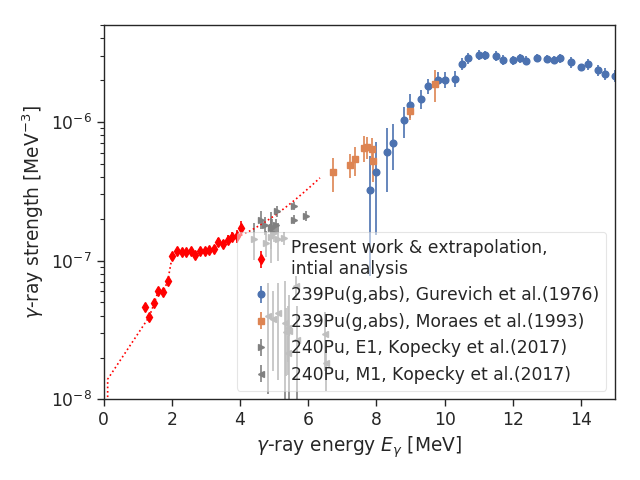}
		\caption{}
	\end{subfigure}
\caption{Initially extracted total level density (a) and $\gamma$SF (b) for $^{240}\mathrm{Pu}$ (a). We used a constant temperature interpolation with $T_\mathrm{CT}$ = 0.415(10). The $\gamma$SF is displayed together with data from~\cite{Kopecky2017, Moraes1993, Gurevich1976} (b). The presented error bars include contributions from both statistical and systematic errors of the unfolding and first generation method~\cite{Schiller2000}.}
\label{fig:ndl_gsf_exp}
\end{figure}

The level density at low $E_x$ was normalized to the discrete level scheme~\cite{NNDC} up the excitation energy where we expect that the low-lying level scheme is known completely ($\approx$ 1\,MeV). At the neutron separation energy $S_\mathrm{n}$, we obtain $\rho(S_n)$ from the average neutron resonance spacing for s-waves, $D_0= 2.20(9)\,\mathrm{eV}$, taken from RIPL-3~\cite{RIPL3} following~\cite{Schiller2000}. The latter conversion depends on the spin-parity distribution; we assumed equal parities and used the spin distribution $g(E_\mathrm{x},I)$  proposed by Ericson Eq. (3.29) ~\cite{Ericson1960} together with the rigid-body moment of inertia approach for the spin cut-off parameter $\sigma$ by von Egidy and Bucurescu (2005)~\cite{Egidy2005}. 
Additionally, we extrapolated from the highest $E_\mathrm{x}$ data points up to $S_\mathrm{n}$. In accordance with findings for other actinides~\cite{Guttormsen2013}, this was performed assuming a constant temperature level density formula~\cite{Gilbert1965}.
The resulting level density $\rho$ is displayed in Fig. \ref{fig:ndl_gsf_exp}a.

The remaining parameter $B$ for the normalization of the transmission coefficient $\mathscr{T}$ can be determined~\cite{Voinov2001, Kopecky1990} from the average total radiative width $\langle \Gamma_\gamma \rangle(S_\mathrm{n}) = 43(4)$~\cite{RIPL3}. The $\gamma$-ray strength function $f(E_\gamma)$ was obtained from the transmission coefficient $\mathscr{T}$ assuming dominance of dipole strength, $f(E_\gamma) = \mathscr{T}(E_\gamma) / (2 \pi E_\gamma^{3})$, and is shown in \ref{fig:ndl_gsf_exp}b.

\section{Impact of the Spin Distribution}

In order to analyze the possible impact of a mismatch between the NLD populated in the $(d,p)$ reaction, $\rho_{pop}$, and the intrinsic NLD, $\rho_{int}$, we will follow a 4-step procedure: 1) identify the \textit{correct} spin distributions $g_\mathrm{pop}$ and $g_\mathrm{int}$, 2) generate synthetic decay data with known NLD and $\gamma$SF, and the identified spin distributions 3) analyze the results with the Oslo Method, and 4) compare the extracted NLD and $\gamma$SF to the input function to infer any systematic deviation.

The (d,p) reaction with the beam energy used in this experiment can be modeled as breakup of a deuteron with emission of a proton, followed by the
formation of a compound nucleus with the remaining
neutron and the target. The spin-parity distribution, $g_\mathrm{pop}(E_\mathrm{x},J,\pi)$, has been calculated in this framework, using the distorted-wave Born approximation (DWBA) in prior form~\cite{Potel2015, Potel2017}. Here we have taken into  account detection angles for the protons and modeled the neutron-nucleus interactions by the dispersive optical model potential of \cite{Capote2008} implemented through potential nr. 2408 listed in \cite{RIPL3}.

\begin{figure}[t]
	\centering
	\includegraphics[width=0.6\linewidth]{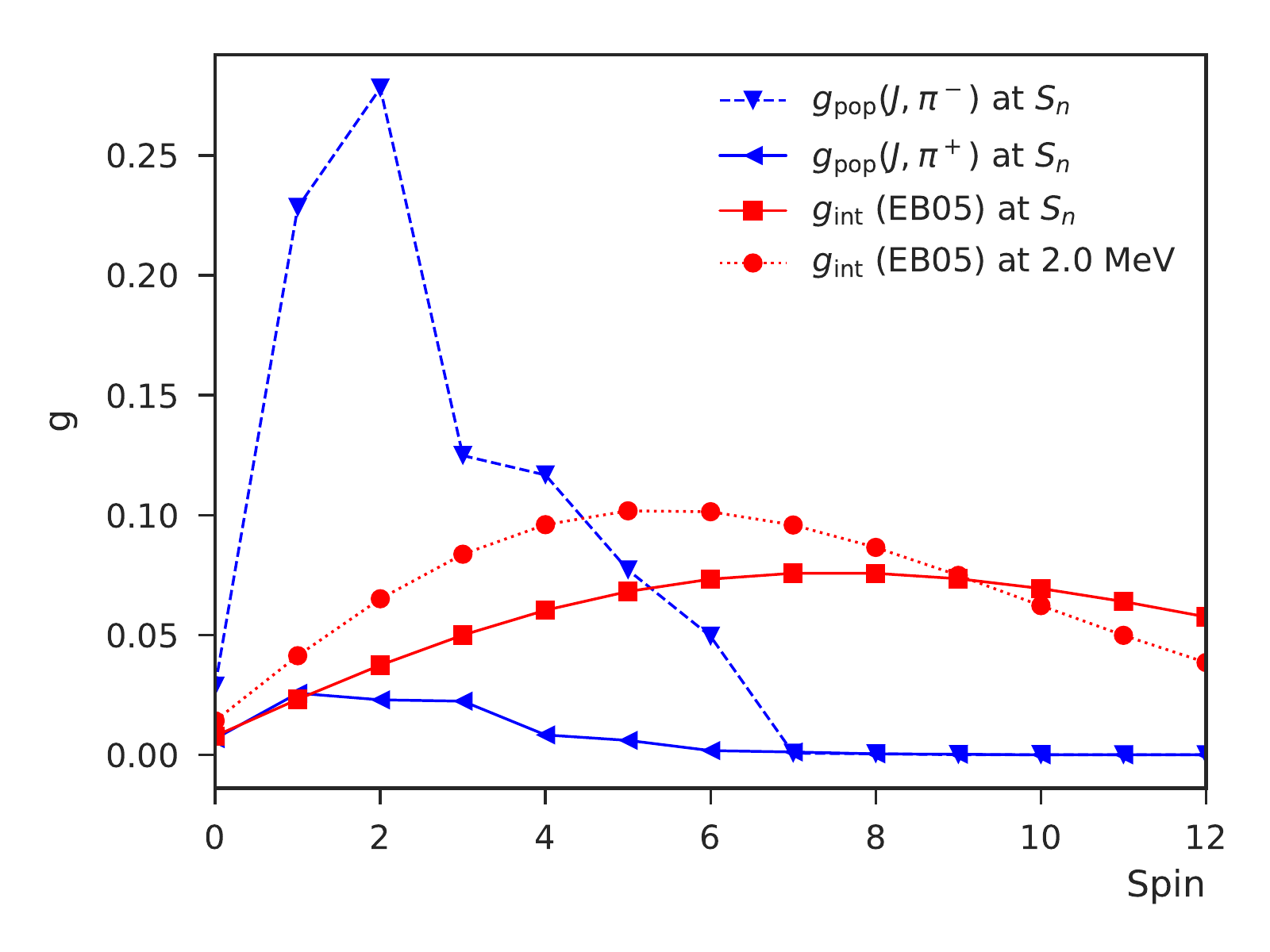}
	\caption{Spin-parity distribution at $E_x = S_n$ of the populated levels $g_\mathrm{pop}$ compared to the intrinsic distribution $g_\mathrm{int}$ at $S_n$. For the latter equiparity is assumed and we also display the distribution at 2 MeV.}
	\label{fig:popToTot}
\end{figure}

To study the effect on the Oslo Method, we first generated a synthetic coincidence data set with the statistical nuclear decay code RAINIER~\cite{Kirsch2018} resembling the (d,p)$^{240}\mathrm{Pu}$ experiment. Following the experimental analysis above, we combined the spin cut-off parameter, $\sigma$, of von Egidy and Bucurescu (2005)~\cite{Egidy2005} with the distribution of 
Ericson~\cite{Ericson1960} to obtain the intrinsic spin-distribution, $g_\mathrm{int}$. As shown in Fig. \ref{fig:popToTot}, the distribution calculated with DWBA of populated levels (further labeled as $g_\mathrm{pop} \neq g_\mathrm{int}$) are centered at much lower spins compared to the assumed intrinsic distribution.

\begin{figure}[t]
	\centering
	\begin{subfigure}[b]{0.49\textwidth}
		\includegraphics[width=\linewidth]{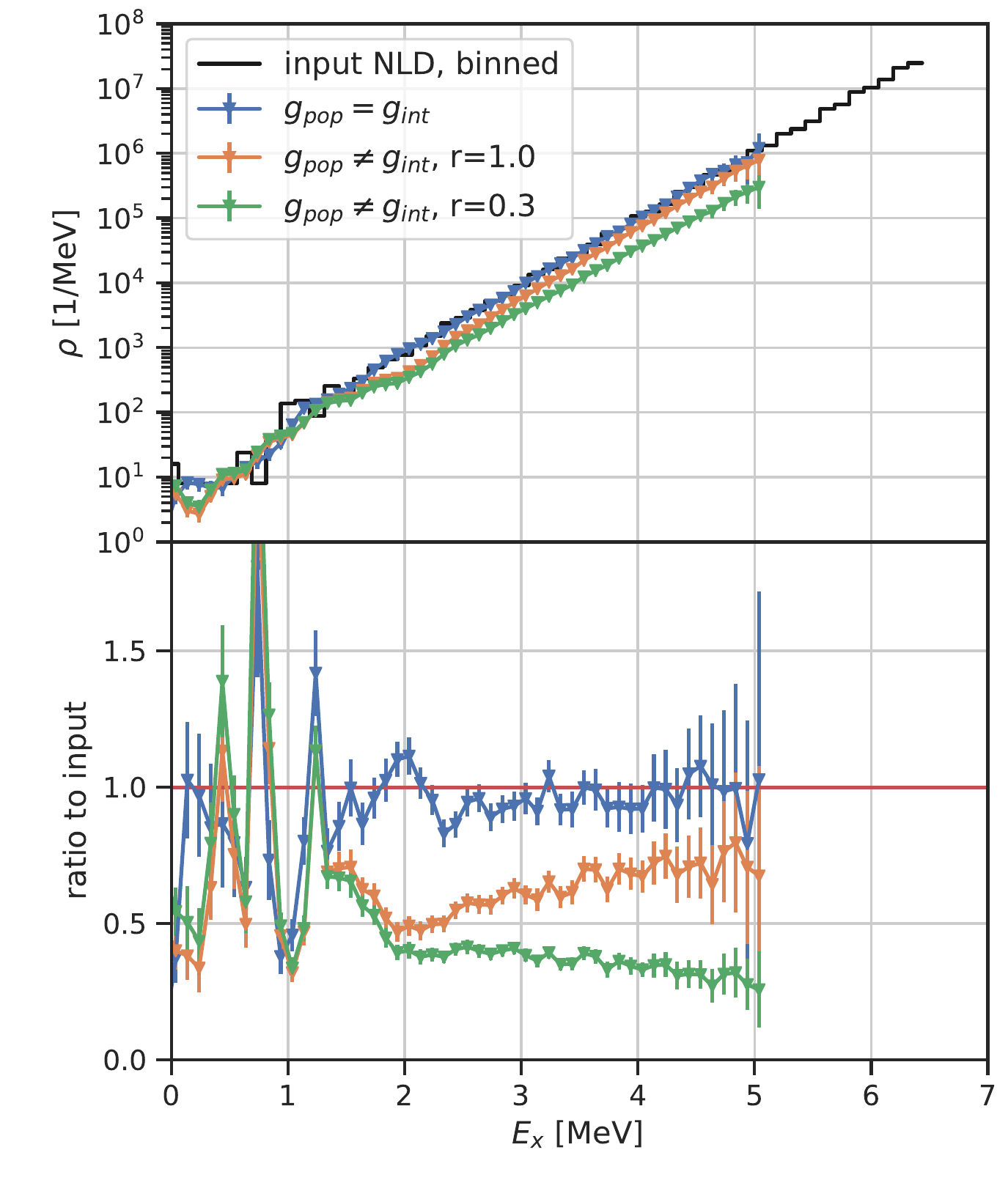}
		\caption{}
		\label{fig:nld_compare}
	\end{subfigure}
	\begin{subfigure}[b]{0.49\textwidth}
		\includegraphics[width=\linewidth]{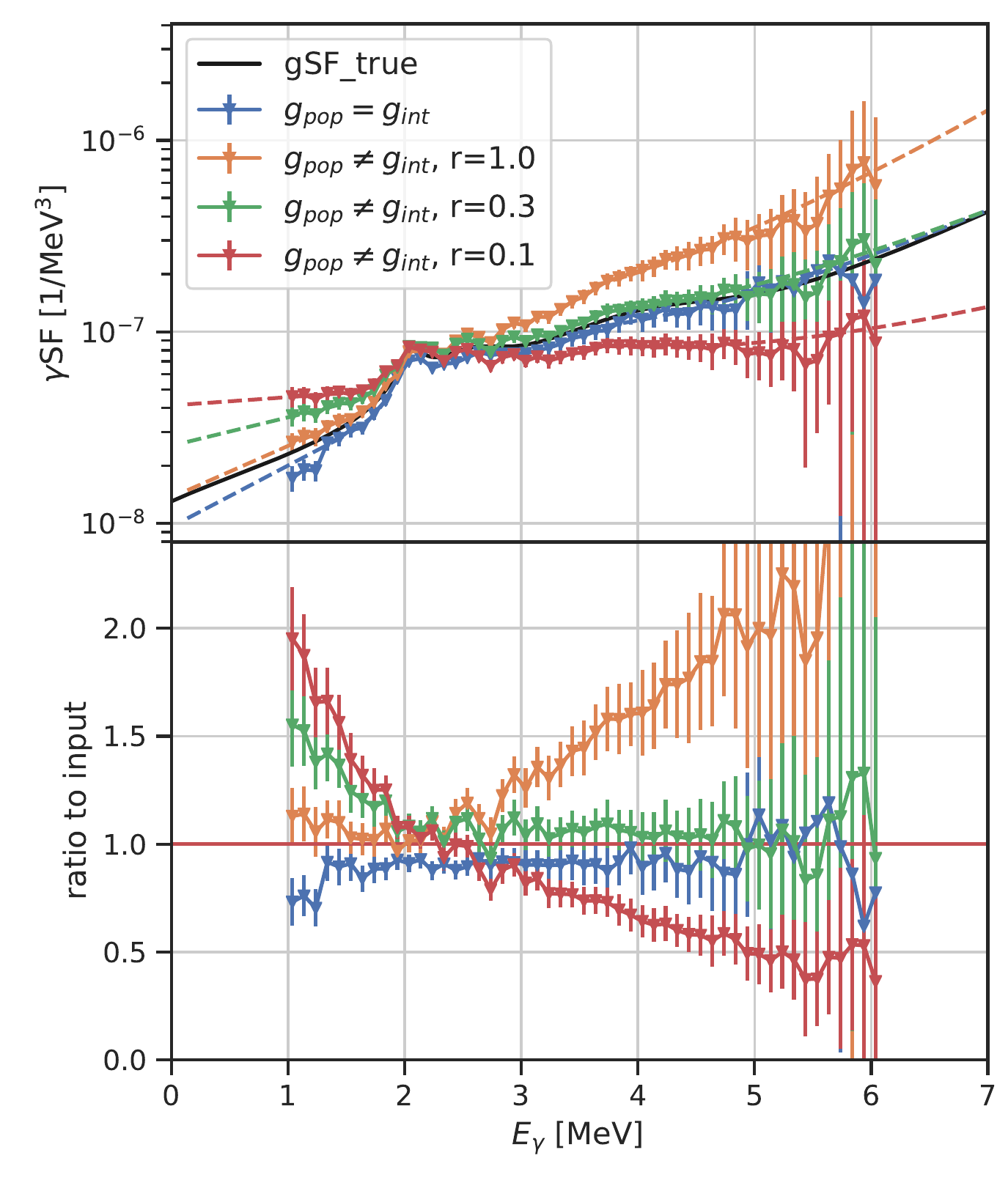}
		\caption{}
		\label{fig:gsf_compare}
	\end{subfigure}
	\caption{Upper panels: NLD (a) and $\gamma$SF (b) extracted with the Oslo Method (with an optional correction $r$) from synthetic dataset. The populated spin distribution $g_\mathrm{pop}$ was either chosen equal to the intrinsic distribution $g_\mathrm{int}$, or narrower, according to the calculations for the (d,p) reaction. Lower panels: Ratio to the known input.}
	\label{fig:ndl_gsf_compare}
\end{figure}

The generated spectra were analyzed with the Oslo Method including folding, unfolding, and the first generation method. The upper panel of Figure \ref{fig:ndl_gsf_compare}a shows the extracted and normalized NLD together with the NLD used as input to RAINIER. The lower panel displays the derived ratio to the input NLD. As expected, it was observed that the NLD in the quasi-continuum (i.e, above the discrete levels) was well reproduced when  $g_\mathrm{pop}=g_\mathrm{int}$; the assumptions of the first-generation method are fulfilled. However, when populating the nucleus by the more realistic, but \textit{narrower} distribution  $g_\mathrm{pop} \neq g_\mathrm{int}$, we underestimated the derived NLD in the quasi-continuum by up to approximately 40\% at 2 MeV. 

This deviation may be qualitatively explained by the smaller fraction of levels populated when decaying with a distribution $g_\mathrm{pop}$ much \textit{narrower} than $g_\mathrm{int}$ (see also Fig. \ref{fig:popToTot}). At higher excitation energies, the ratio is forced to converge to unity due to the normalization at $S_n$.
Note that for the normalization of the $\gamma$SF specified in the next paragraphs, we also display the NLD with $g_\mathrm{pop} \neq g_\mathrm{int}$ where the upper normalization point  $\rho_\mathrm{tot}(S_n)$ obtained from Eq. (28) in~\cite{Schiller2000} was reduced by 
\begin{align}
\rho_\mathrm{red}(S_n) = r  \rho_\mathrm{tot}(S_n), \quad r \leq 1.
\label{eq:rhoRed}
\end{align}
We now turn to the extraction of the $\gamma$SF. For $g_\mathrm{pop}=g_\mathrm{int}$, we observed about $\lesssim 10\%$ difference between the absolute values of the extracted strength and the input function. This difference is mainly attributed to a small mismatch of the true and best-fit temperature for the NLD, which propagates to the $\gamma$SF absolute values through the normalization.

For the more realistic spin distribution $g_\mathrm{pop} \neq g_\mathrm{int}$, we first naively extracted the $\gamma$SF assuming that we had populated all intrinsic levels. Here the shape of the NLD curve is off since it is forced to match the calibration point at $S_\mathrm{n}$. Figure \ref{fig:ndl_gsf_compare}b compares the results to the input $\gamma$SF and although the general shape is preserved, both the slope and absolute value are considerably off as compared to the input. 

Next, we applied a correction inspired by~\cite{Guttormsen2012}, which is based on the assumption that the transmission coefficient $\mathscr{T}$ is spin independent. The first generation matrix, $P$, should then be fit by $P \propto \rho_\mathrm{red} \mathscr{T}$ to extract the correct $\mathscr{T}$, where $\rho_\mathrm{red}$ is obtained from Eq. \eqref{eq:rhoRed} and indicates that we can decay only to a fraction of all intrinsic levels. This will affect the common transformation parameter $\alpha$, see Eq. (\ref{eq:rho_trans},\ref{eq:t_trans}), which determines the slope of the $\gamma$SF: The smaller $r$, the smaller $\alpha$, which translates to a flatter slope.

The determination of the remaining scaling parameter $B$ depends on $\mathscr{T}$ as extracted with $\rho_\mathrm{red}$. However, the level density available for $\gamma$ decay following neutron capture is not effected by this reduction, thus we used $\rho_\mathrm{tot}$ in the $\langle \Gamma_\gamma \rangle$ normalization integral.

We varied the correction factor $r$ and found that $r=0.3$ matches the slope of the input $\gamma$SF best and leads to a constant off-set of about 5-10\%. The larger deviation towards lower $\gamma$-ray energies was traced back to a failure of the first-generation method.

\section{Summary and Conclusions}
We have presented the first systematic analysis of the effect of a realistic, very narrow spin distribution on the Oslo Method for the (d,p) reaction using the example of the heavy nucleus $^{240}\mathrm{Pu}$. We have shown that if the assumptions of the Oslo Method were fulfilled, i.e., if the reaction populated all levels proportionally to the intrinsic spin(-parity) distribution, we regain the correct level density and $\gamma$-ray strength. However, for such a heavy nucleus and a beam energy below the Coulomb barrier, the calculations show a rather small overlap between the populated spins and the intrinsic distribution. This leads to significant distortions in the extracted nuclear level density and $\gamma$-ray strength. We now investigate how the presented approach can be used to correct for the deviations. Finally, the impact on lower mass nuclei needs to be studied, although a significantly greater overlap of the populated and intrinsic distribution and therefore smaller impact is expected.

\section{Acknowledgements}
We would like to thank J.~C.~Müller, E.~A.~Olsen, A.~Semchenkov, and J.~C.~Wikne at the Oslo Cyclotron Laboratory for
providing the stable and high-quality deuterium beam during
the experiment.
This work was supported by the Research Council of Norway under project Grants No. 263030 (F.Z, A.G, S.S) and 262952 (G.M.T), and by the National Research Foundation of South Africa (M.W.). A.C.L. gratefully acknowledges funding from the European Research Council, ERC-STG-2014 Grant Agreement No. 637686. 
This work was performed under the auspices of U.S. Department of Energy by Lawrence Livermore National Laboratory under Contract No. DE-AC52-07NA27344 (D.L.B) and the Lawrence Berkeley National Laboratory under Contract No. DE-AC02-05CH11231 (L.A.B.). The work of B.L.G., J.A.B. and T.A.L. was supported by the U.S. Department of Energy, National Nuclear Security Administration, Office of Defense Nuclear Nonproliferation Research and Development (DNN R\&D) through the Nuclear Science and Security Consortium under Award DE-NA0003180. A.V. was supported under Department of Energy Contract No. DE-\linebreak NA0002905.

%
%
\bibliography{thisbib} 
\end{document}